\documentclass[
reprint,           
superscriptaddress,
amsmath,           
amssymb,           
aps,               
prl,               
notitlepage,       
longbibliography,  
floatfix,          
nofootinbib,
]{revtex4-1}

\usepackage{tensor}     
\usepackage{graphicx}   
\usepackage[
colorlinks=true,        
citecolor=blue,         
linkcolor=blue,         
urlcolor=blue           
]{hyperref}             
\usepackage{bm}         
\usepackage{xcolor}     
\usepackage{lipsum}
\usepackage{color}      
\usepackage[utf8]{inputenc} 
\usepackage[section]{placeins} 
\usepackage{multirow}

\newcommand{\nc}{\newcommand*} 

\nc{\al}{\alpha}
\nc{\s}{\sigma}
\nc{\kp}{\kappa}
\nc{\dt}{\delta}
\nc{\Dt}{\Delta}
\nc{\Ld}{\Lambda}
\nc{\p}{\partial}
\nc{\Gm}{\Gamma}
\nc{\om}{\omega}
\nc{\Om}{\Omega}
\nc{\rd}{\mathrm{d}}
\def\({\left(}
\def\){\right)}
\def\[{\left[}
\def\]{\right]}
\def\e{\begin{equation}}
\def\q{\end{equation}}
\def\m{\begin{eqnarray}}
\def\n{\end{eqnarray}}
\nc{\Eq}[1]{Eq.~\eqref{#1}}     
\nc{\Fig}[1]{Fig.~\ref{#1}}     
\nc{\Table}[1]{Table~\ref{#1}}  
\nc{\Sec}[1]{Sec.~\ref{#1}}     
\nc{\Msun}{M_\odot}             
\nc{\fpbh}{f_{\mathrm{pbh}}}    
\nc{\fpbhn}{f_{\mathrm{pbh0}}}    
\nc{\mR}{\mathcal{R}} 
\nc{\seq}{\sigma_{\mathrm{eq}}}
\nc{\ogw}{\Omega_{\mathrm{GW}}}
\nc{\gpcyr}{\mathrm{Gpc}^{-3}\,\mathrm{yr}^{-1}}
\nc{\lvc}{LIGO/Virgo} 
\nc{\SNR}{\mathrm{SNR}} 
\nc{\mmin}{{m_{\mathrm{min}}}}
\nc{\mmax}{{m_{\mathrm{max}}}}
\nc{\Mmin}{{M_{\mathrm{min}}}}
\nc{\fmin}{{f_{\mathrm{min}}}}
\nc{\VT}{\mathrm{VT}}
\nc{\rhoGW}{\rho_{\mathrm{GW}}}
\nc{\vth}{\vec{\theta}}
\nc{\vd}{\vec{d}}
\nc{\vla}{\vec{\lambda}}
\nc{\Nobs}{N_{\mathrm{obs}}}
\nc{\av}[1]{\langle #1 \rangle} 
\nc{\km}{\mathrm{km}}
\nc{\Mpc}{\mathrm{Mpc}}
\nc{\Tobs}{T_{\mathrm{obs}}}
\nc{\Ntemp}{N_{\mathrm{temp}}}
\nc{\fyr}{f_{\mathrm{yr}}}
\nc{\addref}{[\textcolor{red}{add ref}] } 
\nc{\eg}{\textit{e.g.~}}
\nc{\app}{\approx}
\nc{\hf}{\frac{1}{2}}
\nc{\discuss}{\textcolor{red}{Add discussion here!}}
\nc{\red}[1]{\textcolor{red}{#1}}
\nc{\hp}{h_+} 
\nc{\hc}{h_{\times}} 
\nc{\Oh}{\hat{\Omega}}
\nc{\vx}{\vec{x}}
\nc{\mh}{\hat{m}}
\nc{\nh}{\hat{n}}
\nc{\zh}{\hat{z}}
\nc{\ph}{\hat{p}}
\nc{\A}[1]{\mathcal{A}_{#1}}
\nc{\Ogw}[1]{\Omega_{\mathrm{#1}}}
\nc{\bn}[1]{\dt\bm{t}_{\text{#1}}}
\nc{\bC}[1]{\bm{C}_{\text{#1}}}
\nc{\NTOA}{N_{\text{TOA}}}
\nc{\Nmode}{{N_{\text{mode}}}}
\nc{\ARN}{A_{\rm{RN}}}
\nc{\gRN}{\gamma_{\rm{RN}}}
\nc{\bS}{\mathbf{\Sigma}}
\nc{\br}{\mathbf{r}}
\nc{\bN}{\mathbf{R}}
\nc{\Agw}{A_\mathrm{GWB}}
\nc{\UCP}{\mathrm{UCP}}
\nc{\TT}{\mathrm{TT}}
\nc{\ST}{\mathrm{ST}}
\nc{\SL}{\mathrm{SL}}
\nc{\VL}{\mathrm{VL}}

\nc{\BFST}{$107 \pm 7$}
\begin{document}
	
\title{Non-tensorial Gravitational Wave Background in NANOGrav 12.5-Year Data Set}
	
\author{Zu-Cheng Chen}
\email{chenzucheng@itp.ac.cn} 
\affiliation{CAS Key Laboratory of Theoretical Physics, 
	Institute of Theoretical Physics, Chinese Academy of Sciences,
	Beijing 100190, China}
\affiliation{School of Physical Sciences, 
	University of Chinese Academy of Sciences, 
	No. 19A Yuquan Road, Beijing 100049, China}

\author{Chen Yuan}
\email{yuanchen@itp.ac.cn}
\affiliation{CAS Key Laboratory of Theoretical Physics, 
	Institute of Theoretical Physics, Chinese Academy of Sciences,
	Beijing 100190, China}
\affiliation{School of Physical Sciences, 
	University of Chinese Academy of Sciences, 
	No. 19A Yuquan Road, Beijing 100049, China}

\author{Qing-Guo Huang}
\email{Corresponding author: huangqg@itp.ac.cn}
\affiliation{CAS Key Laboratory of Theoretical Physics, 
Institute of Theoretical Physics, Chinese Academy of Sciences,
Beijing 100190, China}
\affiliation{School of Physical Sciences, 
University of Chinese Academy of Sciences, 
No. 19A Yuquan Road, Beijing 100049, China}
\affiliation{School of Fundamental Physics and Mathematical Sciences
Hangzhou Institute for Advanced Study, UCAS, Hangzhou 310024, China}
\affiliation{Center for Gravitation and Cosmology, 
College of Physical Science and Technology, 
Yangzhou University, Yangzhou 225009, China}

\date{\today}

\begin{abstract}
We perform the first search for an isotropic non-tensorial gravitational-wave background (GWB) allowed in general metric theories of gravity in the North American Nanohertz Observatory for Gravitational Waves (NANOGrav) 12.5-year data set. By modeling the GWB as a power-law spectrum, we find strong Bayesian indication for a spatially correlated process with scalar transverse (ST) correlations whose Bayes factor versus the spatially uncorrelated common-spectrum process is $107\pm 7$, but no statistically significant evidence for the tensor transverse, vector longitudinal and scalar longitudinal polarization modes. The median and the $90\%$ equal-tail amplitudes of ST mode are $\mathcal{A}_{\mathrm{ST}}= 1.06^{+0.35}_{-0.28} \times 10^{-15}$, or equivalently the energy density parameter per logarithm frequency is $\Omega_{\mathrm{GW}}^{\mathrm{ST}} = 1.54^{+1.21}_{-0.71} \times 10^{-9}$, at frequency of 1/year.


\end{abstract}

\maketitle

\textit{Introduction.} The direct detection of gravitational waves (GWs) from compact binary coalescences \cite{TheLIGOScientific:2016pea,LIGOScientific:2018mvr,Abbott:2020niy} has marked the beginning of a new era of GW astronomy and provides a powerful tool to test gravitational physics in the strong-field regime \cite{LIGOScientific:2019fpa,Abbott:2020jks}. The current ground-based GW detectors are sensitive to GWs at frequencies of $10\sim10^4$ Hz \cite{Martynov:2016fzi}. As a complementary tool, the stable millisecond pulsars are natural galactic scale GW detectors that are sensitive in nano-Hertz frequency band, opening  a new window to explore the Universe.
By monitoring the spatially correlated fluctuations induced by GWs on the time of arrivals (TOAs) of radio pulses from an array of pulsars \cite{1978SvA....22...36S,Detweiler:1979wn,1990ApJ...361..300F}, a pulsar timing array (PTA) seeks to detect the very low frequency GWs which might be sourced by the inspiral of supermassive black hole binaries (SMBHBs) \cite{Jaffe:2002rt,Sesana:2008mz,Sesana:2008xk}, the first-order phase transition \cite{Witten:1984rs,Hogan:1986qda}, the scalar-induced GWs \cite{Saito:2008jc,Yuan:2019udt,Yuan:2019wwo}, etc.
The null-detection of GWs by PTAs has successfully constrained various astrophysical scenarios, such as cosmic strings \cite{Lentati:2015qwp,Arzoumanian:2018saf,Yonemaru:2020bmr}, continuous GWs from individual SMBHBs \cite{Zhu:2014rta,Babak:2015lua,Aggarwal:2018mgp}, GW memory effects \cite{Wang:2014zls,Aggarwal:2019ypr}, primordial black holes \cite{Chen:2019xse}, and stochastic GW backgrounds (GWBs) of a power-law spectrum \cite{Lentati:2015qwp,Shannon:2015ect,Arzoumanian:2018saf}. However, the direct detection of GWs by PTAs remains a key task in astrophysical experiments, and is hopefully achieved in the next few years \cite{Siemens:2013zla,Taylor:2015msb}.
 
Recently, the North American Nanohertz Observatory for Gravitational Waves (NANOGrav) collaboration has reported strong evidence for a stochastic common-spectrum process, which is significantly preferred over an independent red-noise process in each pulsar \cite{Arzoumanian:2020vkk}. The characteristic strain of this process is described by a power-law model, $h_c(f)\propto f^{-2/3}$, corresponding to the GW emission from inspiraling SMBHBs.
NANOGrav announced there was no statistically significant evidence for quadrupolar spatial correlations. Moreover, this process shows moderately negative evidence for monopolar and dipolar correlations, which may come from the reference clock and solar system ephemeris (SSE) anomalies, respectively. Lacking definitive evidence for quadrupolar spatial correlations \cite{Arzoumanian:2020vkk}, NANOGrav argued that it is inconclusive to claim a detection of GWB consistent with general relativity (GR), and the origin of this process remains controversial.

Even though there is no definitive evidence for tensor transverse (TT) correlations predicted by GR in the NANOGrav 12.5-year data set, it does not exclude the possibility of other GW polarization modes allowed in general metric theories of gravity. In fact, a most general metric gravity theory can allow two vector modes and two scalar modes besides the two tensor modes, and these different modes have distinct correlation patterns \cite{2008ApJ...685.1304L,Chamberlin:2011ev,Gair:2015hra,Boitier:2020xfx}, allowing the GW detectors to explore them separately. To figure out whether the signal originates from a GWB or not, it is necessary to fit the data with all possible correlation patterns. In this letter, we perform the first Bayesian search for the stochastic GWB signal modeled by a power-law spectrum with all the six polarization modes in the NANOGrav 12.5-year data set. Such a power-law spectrum of GWB can be produced by the inspiraling SMBHBs by assuming circular orbits whose decays are dominated by GWs and neglecting higher moments \cite{Cornish:2017oic}. We find the Bayes factor in favor of a spatially correlated common-spectrum process with the scalar transverse (ST) correlations versus the spatially uncorrelated common-spectrum process (UCP) is {\BFST} which indicates that strong Bayesian indication for the ST correlations in the NANOGrav 12.5-year data set.



\textit{Detecting GWB Polarizations with a PTA.}
The radio pulses from pulsars, especially millisecond pulsars, arrive at the Earth at extremely steady rates, and pulsar timing experiments exploit this regularity. The geodesics of the radio waves can be perturbed by GWs, inducing the fluctuations in the TOAs of radio pulses \cite{1978SvA....22...36S,Detweiler:1979wn}. The presence of a GW will manifest as the unexplained residuals in the TOAs after subtracting a deterministic timing model that accounts for the pulsar spin behavior and the geometric effects due to the motion of the pulsar and the Earth \cite{1978SvA....22...36S,Detweiler:1979wn}. By regularly monitoring TOAs of pulsars from an array of the ultra rotational stable millisecond pulsars \cite{1990ApJ...361..300F} and using the expected form for cross correlations of a signal between pulsars in the array, it is feasible to discriminate the GW signal from other systematic effects, such as clock or SSE errors.

For any two pulsars ($a$ and $b$) in a PTA, the cross-power spectral density of the timing residuals induced by a GWB at frequency $f$ will be \cite{2008ApJ...685.1304L,Chamberlin:2011ev,Gair:2015hra}
\e\label{Sab1}
S_{ab}(f) = \sum_P \frac{h_{c,P}^2}{12 \pi^2 f^3} \Gm^P_{ab}(f),
\q 
where $h_c^P(f)$ is the characteristic strain and the sum is over all the six possible GW polarizations which may be presented in a general metric gravity theory, namely $P = +, \times, x, y, l, b$. Here, ``$+$" and ``$\times$" denote the two different spin-2 transverse traceless polarization modes; ``$x$" and ``$y$" denote the two spin-1 shear modes; ``$l$" denotes the spin-0 longitudinal mode; and ``$b$" denotes the spin-0 breathing mode. The overlap function $\Gm^P_{ab}$ for two pulsars is given by \cite{2008ApJ...685.1304L,Chamberlin:2011ev}
\m
\Gm^P_{ab}(f) =&&\frac{3}{8\pi} \int d\Oh \(e^{2\pi i f L_a(1+\Oh\cdot\ph_a)}-1\)\times\nonumber\\
&&\(e^{2\pi i f L_b(1+\Oh\cdot\ph_b)}-1\) F^P_a(\Oh) F^P_b(\Oh),
\n 
where $L_a$ and $L_b$ are the distance from the Earth to the pulsar $a$ and $b$ respectively, $\Oh$ is the propagating direction of the GW, and $\ph$ is the direction of the pulsar with respect to the Earth. The antenna patterns $F^P(\Oh)$ are given by
\e 
F^P(\Oh) = e^P_{ij}(\Oh) \frac{\ph^i\ph^j}{2(1+\Oh\cdot\ph)},
\q
where $e^P_{ij}$ is the polarization tensor for polarization mode $P$ \cite{2008ApJ...685.1304L,Chamberlin:2011ev}. Following \cite{Cornish:2017oic}, we define
\m 
\Gm^{\TT}_{ab}(f) &=& \Gm^{+}_{ab}(f) + \Gm^{\times}_{ab}(f), \label{gammaTT}\\
\Gm^{\ST}_{ab}(f) &=& \Gm^{b}_{ab}(f), \label{gammaST}\\
\Gm^{\VL}_{ab}(f) &=& \Gm^{x}_{ab}(f) + \Gm^{y}_{ab}(f), \label{gammaVL}\\
\Gm^{\SL}_{ab}(f) &=& \Gm^{l}_{ab}(f). \label{gammaSL}
\n 
For the $\TT$ and $\ST$ polarization modes, the overlap functions are approximately independent of the distance and frequency and can be analytically calculated by \cite{Hellings:1983fr,2008ApJ...685.1304L}
\m\label{TTST}
\Gm^{\TT}_{ab}(f) &=& \hf(1+\dt_{ab})+ \frac{3}{2} k_{ab} \(\ln k_{ab}-\frac{1}{6}\), \\
\Gm^{\ST}_{ab}(f) &=& \frac{1}{8}\(3 + 4\dt_{ab} + \cos\zeta_{ab}\),
\n 
where $\dt_{ab}$ is the Kronecker delta symbol, $\zeta_{ab}$ is the angle between pulsars $a$ and $b$, and $k_{ab} \equiv (1-\cos\zeta_{ab})/2$. Note that $\Gm^{\TT}_{ab}$ is known as the Hellings \& Downs (HD) \cite{Hellings:1983fr} or quadrupolar correlations. However, there exist no analytical expressions for the vector longitudinal ($\VL$) and scalar longitudinal ($\SL$) polarization modes, and we calculate them numerically.

PTAs are sensitive to the GWs at frequencies of approximately $10^{-9}\sim10^{-7}$ Hz, and it is expected that the GWB from a population of inspiraling SMBHBs will be the dominant source in this frequency band \cite{Jaffe:2002rt,Sesana:2008mz,Sesana:2008xk}. Assuming the binaries are in circular orbits and the orbital decay is dominated by the GW emission, the cross-power spectral density of \Eq{Sab1} can be approximately estimated by \cite{Cornish:2017oic}
\e\label{Sab2}
    S_{ab}(f) = \sum_{I={\TT, \ST, \VL, \SL}} \Gm^I_{ab} \frac{\A{I}^2}{12\pi^2}   \(\frac{f}{\fyr}\)^{-\gamma_I} \fyr^{-3},
\q
where $\A{I}$ is the GWB amplitude of polarization mode $I$, and $\fyr = 1/\mathrm{year}$. The power-law index $\gamma_I$ for the TT polarization is $\gamma_\TT = 13/3$, and $\gamma_\ST = \gamma_\VL = \gamma_\SL = 5$ for other polarizations. The dimensionless GW energy density parameter per logarithm frequency for the polarization mode $I$ is related to $\A{I}$ by, \cite{Thrane:2013oya}, 
\e 
    \ogw^{I}(f) = \frac{2\pi^2}{3 H_0^2} f^2 h_{c,I}^2 = \frac{2\pi^2\fyr^2}{3 H_0^2} \A{I}^2 \(\frac{f}{\fyr}\)^{5-\gamma_I},
\q 
where $H_0$ is the Hubble constant and we take $H_0 = 67.4\, \km \sec^{-1} \Mpc^{-1}$ from Planck 2018 \citep{Aghanim:2018eyx}.


\textit{PTA data analysis.} The NANOGrav collaboration has searched the isotropic GWB in their 12.5-year timing data set \cite{Alam:2020fjy} and found strong evidence for a stochastic common-spectrum process but without statistically significant evidence for the TT spatial correlations \cite{Arzoumanian:2020vkk}. In this letter, we perform the first search for the GWB from the non-tensorial polarization modes in the NANOGrav 12.5-year data set. 

\begin{table*}[htbp!]
    \scriptsize
    \caption{Parameters and their prior distributions used in the analyses.}
    \label{tab:priors}
    \begin{tabular}{llll}
        \hline\hline
        parameter & description & prior & comments \\
        \hline
        \multicolumn{4}{c}{White Noise} \\[1pt]
        $E_{k}$ & EFAC per backend/receiver system & Uniform $[0, 10]$ & single-pulsar analysis only \\
        $Q_{k}$[s] & EQUAD per backend/receiver system & log-Uniform $[-8.5, -5]$ & single-pulsar analysis only \\
        $J_{k}$[s] & ECORR per backend/receiver system & log-Uniform $[-8.5, -5]$ & single-pulsar analysis only \\
        \hline
        \multicolumn{4}{c}{Red Noise} \\[1pt]
        $\ARN$ & red-noise power-law amplitude & log-Uniform $[-20, -11]$ & one parameter per pulsar  \\
        $\gRN$ & red-noise power-law spectral index & Uniform $[0, 7]$ & one parameter per pulsar \\
        \hline
        \multicolumn{4}{c}{Uncorrelated Common-spectrum Process (UCP)} \\[1pt]
        $\A{\UCP}$ & UCP power-law amplitude & log-Uniform $[-18, -14]$ & one parameter for PTA \\
        $\gamma_{\UCP}$ & UCP power-law spectral index & delta function ($\gamma_{\UCP}=13/3$) & fixed \\
        \hline
        \multicolumn{4}{c}{GWB Process} \\[1pt]
        $\A{\TT}$ & GWB amplitude of TT polarization & log-Uniform $[-18, -14]$ & one parameter for PTA \\
        $\A{\ST}$ & GWB amplitude of ST polarization & log-Uniform $[-18, -14]$ & one parameter for PTA \\
        $\A{\VL}$ & GWB amplitude of VL polarization & log-Uniform $[-19, -15]$ & one parameter for PTA \\
        $\A{\SL}$ & GWB amplitude of SL polarization & log-Uniform $[-20, -16]$ & one parameter for PTA \\
        \hline
        \multicolumn{4}{c}{\textsc{BayesEphem}} \\[1pt]
        $z_{\rm drift}$ [rad/yr] & drift-rate of Earth's orbit about ecliptic $z$-axis & Uniform [$-10^{-9}, 10^{-9}$] & one parameter for PTA \\
        $\Delta M_{\rm jupiter}$ [$M_{\odot}$] & perturbation to Jupiter's mass & $\mathcal{N}(0, 1.55\times 10^{-11})$  & one parameter for PTA \\
        $\Delta M_{\rm saturn}$ [$M_{\odot}$] & perturbation to Saturn's mass & $\mathcal{N}(0, 8.17\times 10^{-12})$  & one parameter for PTA \\
        $\Delta M_{\rm uranus}$ [$M_{\odot}$] & perturbation to Uranus' mass & $\mathcal{N}(0, 5.72\times 10^{-11})$  & one parameter for PTA \\
        $\Delta M_{\rm neptune}$ [$M_{\odot}$] & perturbation to Neptune's mass & $\mathcal{N}(0, 7.96\times 10^{-11})$  & one parameter for PTA \\
        PCA$_{i}$ & principal components of Jupiter's orbit & Uniform $[-0.05, 0.05]$ & six parameters for PTA \\
        \hline
    \end{tabular}
\end{table*}

Following NANOGrav \cite{Arzoumanian:2020vkk}, in our analyses, we use 45 pulsars whose timing baseline is greater than three years. 
To calculate the longitudinal response functions $\Gm^{\VL}_{ab}$ and $\Gm^{\SL}_{ab}$ that are dependent on the pulsar distance from the Earth, we adopt the distance information from the Australia Telescope National Facility (ATNF) pulsar database\footnote{\url{https://www.atnf.csiro.au/research/pulsar/psrcat/}} \cite{Manchester:2004bp}. Due to the uncertainty in the pulsar distance measurement, the estimation uncertainty of the overlap function can be $\lesssim 3\%$ for VL mode, and $\lesssim 20\%$ for SL mode. The timing residuals of each single pulsar after subtracting the timing model from the TOAs can be decomposed as \cite{Arzoumanian:2015liz}
\e\label{dt}
    \dt\bm{t} = M \bm{\epsilon} + F \bm{a} + \bm{n}.
\q
The term $M \bm{\epsilon}$ accounts for the inaccuracies in the subtraction of timing model, where $M$ is the timing model design matrix and $\bm{\epsilon}$ is a vector denoting small offsets for the timing model parameters. The timing model design matrix is obtained through  \texttt{libstempo}\footnote{\url{https://vallis.github.io/libstempo}} package which is a python interface to \texttt{TEMPO2} \footnote{\url{https://bitbucket.org/psrsoft/tempo2.git}} \cite{Hobbs:2006cd,Edwards:2006zg} timing software. The term $F \bm{a}$ describes all low-frequency signals, including both the red noise intrinsic to each pulsar and the common red noise signal common to all pulsars (such as a GWB), where $F$ is the Fourier design matrix with components of alternating sine and cosine functions and $\bm{a}$ is a vector giving the amplitude of the Fourier basis functions at the frequencies of $\{1/T, 2/T, \cdots, \Nmode/T\}$ with $T$ the span between the minimum and maximum TOA in the PTA \cite{vanHaasteren:2014qva}. Similar to NANOGrav \cite{Arzoumanian:2020vkk}, we use $30$ frequency components ($\Nmode=30$) for the pulsar intrinsic red noise with a power-law spectrum while using $5$ frequency components ($\Nmode=5$) for the common-spectrum process to mitigate the effect of potentially coupling between the higher-frequency components of common red noise process and the white noise \cite{Arzoumanian:2020vkk}. The last term $\bm{n}$ describes the timing residuals induced by white noise, including a scale parameter on the TOA uncertainties (EFAC), an added variance (EQUAD), and a per-epoch variance (ECORR) for each backend/receiver system \cite{Arzoumanian:2015liz}.


Similar to NANOGrav \cite{Arzoumanian:2020vkk}, we use the latest JPL SSE, DE438 \cite{DE438}, as the fiducial SSE. For verification, we also allow for the \textsc{BayesEphem} \cite{Vallisneri:2020zhb} corrections to DE438 to model the SSE uncertainties. However, one should bear in mind that introducing \textsc{BayesEphem} would subtract the power from the putative GWB process and suppress the evidence of the GWB process \cite{Vallisneri:2020zhb,Arzoumanian:2020vkk,Pol:2020igl}. To extract information from the data, we perform the Bayesian parameter inferences by closely following the procedure in \cite{Arzoumanian:2018saf,Arzoumanian:2020vkk}. The parameters of our models and their prior distributions are summarized in \Table{tab:priors}. To reduce the computational costs, in our analyses, we fixed the white noise parameters to their max likelihood values from results released by NANOGrav\footnote{\url{https://github.com/nanograv/12p5yr_stochastic_analysis}}. We use \texttt{enterprise} \cite{enterprise} and \texttt{enterprise\_extension}\footnote{\url{https://github.com/nanograv/enterprise_extensions}} software packages to calculate the likelihood and Bayes factors and use \texttt{PTMCMCSampler} \cite{justin_ellis_2017_1037579} package to do the Markov chain Monte Carlo sampling. To reduce the number of samples needed for the chains to burn in, we use draws from empirical distributions to sample the pulsars' red noise parameters as was done in \cite{Aggarwal:2018mgp,Arzoumanian:2020vkk}, with the distributions based on the posteriors obtained from an initial Bayesian analysis that includes only the pulsars' red noise (\textit{i.e.} excluding any common red noise process). 


Our analysis is mainly based on the Bayesian inference in which the Bayes factor $\mathcal{B}_{10}\equiv \rm{Pr}(\mathcal{D}|\mathcal{M}_1)/\rm{Pr}(\mathcal{D}|\mathcal{M}_0)$ is used to quantify the model selection, where $\rm{Pr}(\mathcal{D}|\mathcal{M})$ denotes the probability that the data $\mathcal{D}$ are produced under the assumption of model $\mathcal{M}$. In \cite{BF}, $\mathcal{B}_{10}\in [20,150]$ and $\mathcal{B}_{10}>150$ respectively correspond to strong and very strong evidence for $\mathcal{M}_1$. More optimistically, $\mathcal{B}_{10}\in [10,30]$, $\mathcal{B}_{10}\in [30,100]$, and $\mathcal{B}_{10}>100$ correspond to strong, very strong and extreme evidence for $\mathcal{M}_1$ in \cite{lee_wagenmakers_2014}. NANOGrav found strong evidence for a common-spectrum process in the 12.5-year data set and reported the Bayes factors of UCP model versus the pulsar-intrinsic red noise only model to be $10^{4.5}$ with DE438, and $10^{2.4}$ with \textsc{BayesEphem} \cite{Arzoumanian:2020vkk}. In this letter, the UCP model with fixed spectral index $\gamma_{\UCP}=13/3$ is taken as the fiducial model $\mathcal{M}_0$, and the model $\mathcal{M}_1$ with $\mathcal{B}_{10} \gg 1$ is supposed to be significantly preferred over the UCP model. We perform analyses on various models by considering different correlation combinations as presented in \Eq{Sab2}. 


\begin{table}[htbp!]
    \begin{center}
        \begin{tabular}{c|cccc}
            \hline\hline
            ephemeris & TT &  ST  & VL & SL  \\
            \hline
            DE438& $4.96(9)$ & $107(7)$& $1.94(3)$ & $0.373(5)$  \\
            \hline
            \textsc{BayesEphem} & $2.35(3)$ & $18.4(7)$ &$1.31(2)$ & $0.555(7)$ \\
            \hline
        \end{tabular}
    \end{center}  
    \caption{\label{bayes}The Bayes factors for various models compared to the UCP model with $\gamma=13/3$. The digit in the parentheses gives the uncertainty on the last quoted digit.}
\end{table}

\begin{figure}[htbp!]
    \centering
    \includegraphics[width=\linewidth]{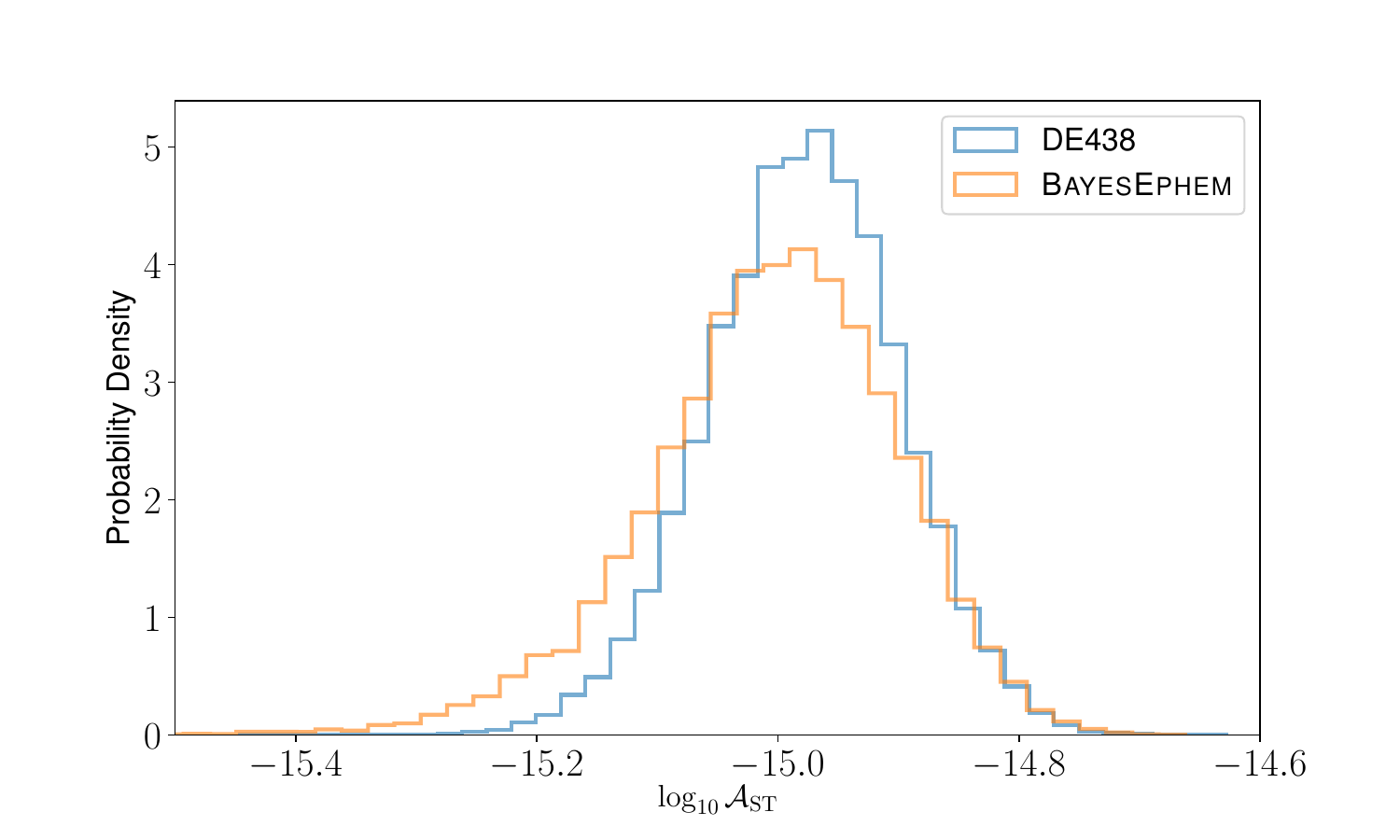}
    \caption{\label{post_ST} Bayesian posteriors for the ST amplitude $\A{\ST}$ obtained in the ST model under DE438 and \textsc{BayesEphem} ephemeris schemes, respectively.}
\end{figure}

\begin{figure}[htbp!]
	\centering
	\includegraphics[width=\linewidth]{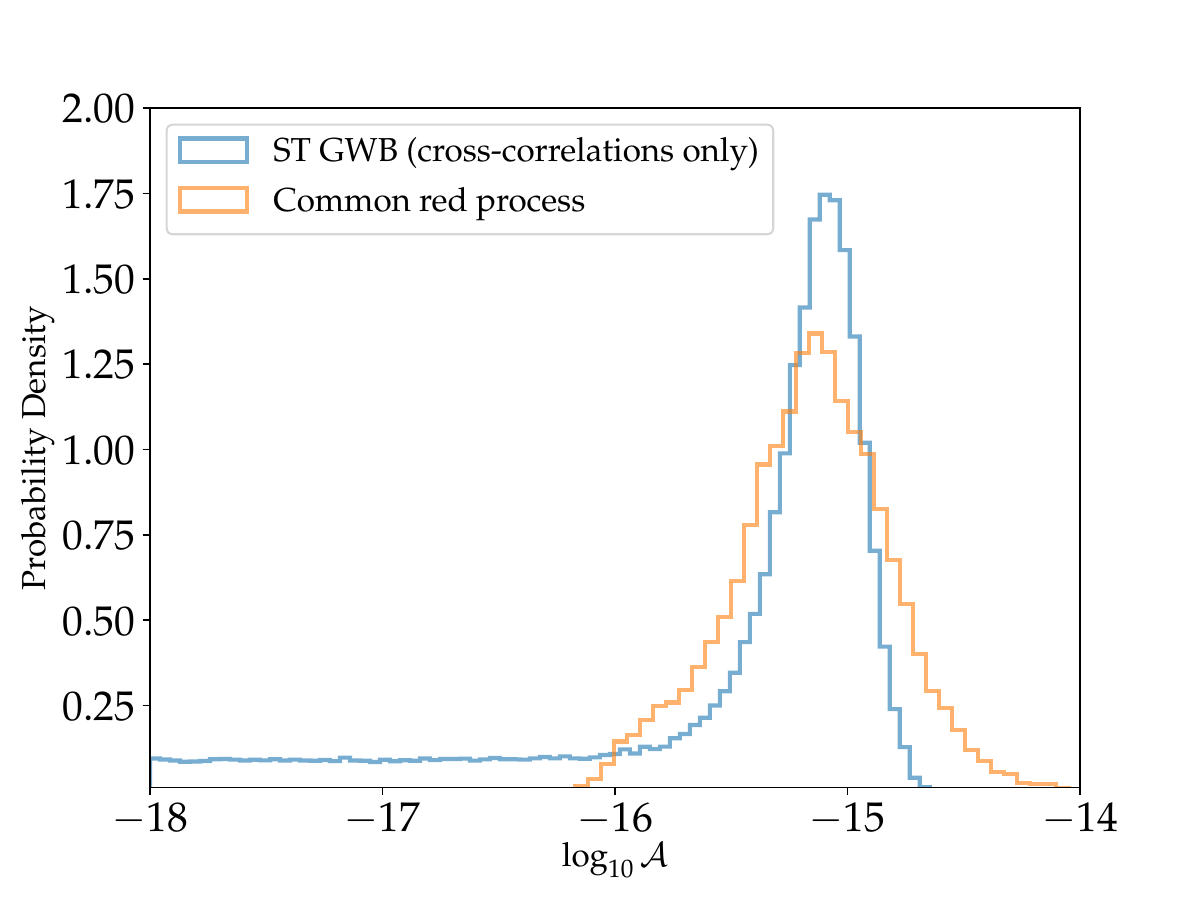}
	\caption{\label{Bayesian_amp} Bayesian amplitude posteriors in a model  (with DE438) that includes a common-spectrum process and an off-diagonal ST-correlated process where all auto-correlation terms are set to zero. The posteriors shown here are marginalized to each other.
	}
\end{figure}

\textit{Results and discussion.} Our results are summarized in \Table{bayes} in which we list the Bayes factors for different models with respect to the UCP model. The Bayes factor of the TT model compared to the UCP model is $4.96\pm 0.09$ with DE438, and $2.35\pm 0.03$ with \textsc{BayesEphem}, indicating no statistically significant evidence for the TT correlations in the data, which is consistent with the results from NANOGrav \cite{Arzoumanian:2020vkk}. The Bayes factors of VL and SL models compared to the UCP model are smaller than $3$, implying the VL and SL signals are ``not worth more than a bare mention" \cite{BF}. However, the Bayes factor for the ST model versus the UCP model is {\BFST} with DE438, implying strong indication for the ST correlations \cite{BF,lee_wagenmakers_2014}, and we obtain the median and the $90\%$ equal-tail amplitudes as $\A{\ST} = 1.06^{+0.35}_{-0.28} \times 10^{-15}$ or equivalently $\ogw^\ST = 1.54^{+1.21}_{-0.71} \times 10^{-9}$, at frequency of 1/year. It is known that \textsc{BayesEphem} may absorb a common-spectrum process and weaken the evidence of the GWB process if it exists in the data \cite{Vallisneri:2020zhb,Arzoumanian:2020vkk,Pol:2020igl}. Nevertheless, even in the case of \textsc{BayesEphem}, the Bayes factor for the ST model is $18.4\pm 0.7$ which is still significant in the sense of statistics. See the Bayesian posteriors for the ST amplitude $\A{\ST}$ obtained in the ST model in \Fig{post_ST}.
Although NANOGrav reported the UCP is more consistent with $\gamma=5.5$ \cite{Arzoumanian:2020vkk}, we found that such a large Bayes factor for ST model versus the UCP model cannot be explained by the ST spectral index $\gamma_{\rm{ST}}=5$ because the Bayes factor for the ST model versus the UCP model with $\gamma=5.5$ is $96 \pm 9$ with DE438. It implies that the preferred ST model is likely attributed to the cross-correlations. Furthermore, we also consider a model that includes a common-spectrum process and an off-diagonal ST-correlated process where all auto-correlation terms are set to zero. The Bayesian amplitude posteriors are shown in Fig.~\ref{Bayesian_amp} in which the amplitude posterior of the off-diagonal ST-correlated process is significant and comparable to the amplitude posterior of the common-spectrum process, indicating that the large Bayes factor for the ST model should be attributed to the cross-correlations in the NANOGrav 12.5-year data set.

\begin{figure}[htbp!]
    \centering
    \includegraphics[width=\linewidth]{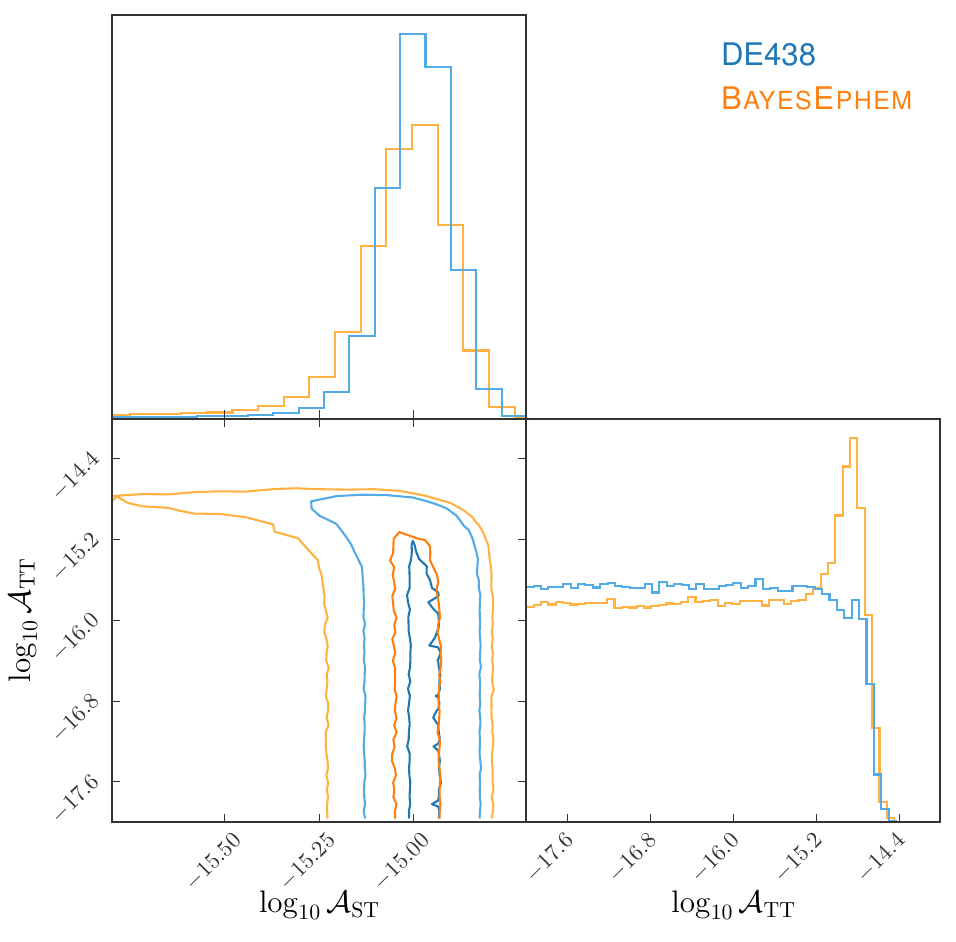}
    \caption{\label{post_ST+TT} One and two-dimensional marginalized posteriors of ST and TT amplitudes obtained from the ST+TT model under DE438 and \textsc{BayesEphem} ephemeris schemes, respectively. We show both the $1 \sigma$ and $2 \sigma$ contours in the two-dimensional plot.}
\end{figure}

In addition, we also consider an ST+TT model in which we simultaneously take into account both the ST and TT correlations. The contour plot and the posterior distributions of the ST and TT amplitudes in the ST+TT model are shown in \Fig{post_ST+TT}, which implies that the presence of ST correlations is preferred even using \textsc{BayesEphem}, but no significant evidence for additional TT correlations. The amplitude of ST mode from this model with DE438 is $\A{\ST} = 1.02^{+0.36}_{-0.44} \times 10^{-15}$ or equivalently $\ogw^\ST = 1.45^{+1.21}_{-0.98} \times 10^{-9}$, at frequency of 1/year. This result is consistent with the former one in the ST model.


To summarize, we find strong Bayesian indication for the ST correlations but no statistically significant evidence for the TT, VL, and SL correlations in the NANOGrav 12.5-year data set. We hope that the future PTA data sets growing in timespan and number of pulsars continue to confirm our results presented in this letter.

{\it Acknowledgments. }
We would like to thank the anonymous referee for the useful suggestions and comments. We also acknowledge the use of HPC Cluster of ITP-CAS and HPC Cluster of Tianhe II in National Supercomputing Center in Guangzhou. This work is supported by the National Key Research and Development Program of China Grant No.2020YFC2201502, grants from NSFC (grant No. 11975019, 11690021, 11991052, 12047503),  Strategic Priority Research Program of Chinese Academy of Sciences (Grant No. XDB23000000, XDA15020701), and Key Research Program of Frontier Sciences, CAS, Grant NO. ZDBS-LY-7009.
	
\bibliography{./ref}

\end{document}